\documentclass{elsart}
\usepackage{epsfig}
\usepackage{amssymb}
\def\Gn{$\Gamma_n$}
\def\Gp{$\Gamma_p$}
\def\Gnp{$\Gamma_n/\Gamma_p$}
\def\G2N{$\Gamma_{2N}$}

\def\He5{$^5$He}
\def\Li6lam{$^{6}_{\Lambda}$Li}
\def\Clam{$^{12}_{\it{\Lambda}}$C}
\def\Li6{$^{6}$Li}
\def\He4lam{$^4_\Lambda$He}
\def\He5lam{$^{5}_{\it{\Lambda}}$He}

\def\fe-lambda{$\  ^ {56}_ \Lambda Fe\ $}

\def\bb{$\it{bb}$~}

\def\nn{$nn$~}
\def\np{$np$~}

\begin{document}
\begin{frontmatter}

\title{Coincidence Measurement of the Nonmesonic Weak Decay of $^{12}_{\it{\Lambda}}$C}
\author[snu]{M. J. Kim},
\ead{mijung@ieplab.snu.ac.kr}
\author[osaka]{S. Ajimura},
\author[kek]{K. Aoki},
\author[gsi]{A. Banu},
\author[snu]{H. Bhang},
\author[OEC]{T. Fukuda},
\author[tohoku]{O. Hashimoto},
\author[snu]{J. I. Hwang},
\author[tohoku]{S. Kameoka},
\author[snu]{B. H. Kang}, 
\author[snu]{E. H. Kim},
\author[snu]{J. H. Kim\thanksref{NOWkim}},
\author[ut]{T. Maruta},
\author[tohoku]{Y. Miura},
\author[osaka]{Y. Miyake},
\author[kek]{T. Nagae},
\author[ut]{M. Nakamura}, 
\author[tohoku]{S. N. Nakamura},
\author[kek]{H. Noumi},
\author[titech]{S. Okada\thanksref{NOWriken}},
\author[tohoku]{Y. Okayasu},
\author[kek]{H. Outa\thanksref{NOWriken}},
\author[kriss]{H. Park},
\author[kek]{P. K. Saha\thanksref{NOWjaeri}},
\author[kek]{Y. Sato}, 
\author[kek]{M. Sekimoto},
\author[ewha]{S. Shin},
\author[tohoku]{T. Takahashi\thanksref{NOWkek}},
\author[tohoku]{H. Tamura},
\author[riken]{K. Tanida},
\author[kek]{A. Toyoda}, 
\author[tohoku]{K. Tsukada},
\author[tohoku]{T. Watanabe},
\author[snu]{H. J. Yim}

\address[snu]{Department of Physics, Seoul National University,
 Seoul 151-742, Korea}
\address[osaka]{Department of Physics, Osaka University,
 Toyonaka 560-0043, Japan}
\address[kek]{High Energy Accelerator Research Organization (KEK),
 Tsukuba 305-0801, Japan}
\address[OEC]{Laboratory of Physics, Osaka Electro Communication 
 University, Neyagawa 572-8530, Japan}
\address[tohoku]{Department of Physics, Tohoku University,
 Sendai 980-8578, Japan}
\address[ut]{Department of Physics, University of Tokyo,
 Hongo 113-0033, Japan}
\address[titech]{Department of Physics, Tokyo Institute of Technology,
 Ookayama 152-8551, Japan}
\address[kriss]{Korea Research Institute of Standards and Science (KRISS),
 Daejeon 305-600, Korea}
\address[gsi]{Gesellschaft f$\ddot{\mbox{u}}$r Schwerionenforschung mbH (GSI),
 Darmstadt 64291, Germany}
\address[ewha]{Department of Physics, Ewha Womans University,
 Seoul 11-1, Korea}
\address[riken]{RIKEN Wako Institute, RIKEN,
 Wako 351-0198, Japan}
\thanks[NOWriken]{Present address: RIKEN Wako Institute, RIKEN,
  Wako 351-0198, Japan}
\thanks[NOWjaeri]{Present address:
 Japan Atomic Energy Agency, Tokai, Ibaraki 319-1195, Japan}
\thanks[NOWkek]{Present address:
 High Energy Accelerator Research Organization (KEK), Tsukuba 305-0801, Japan}
\thanks[NOWkim]{Present address:
 Korea Research Institute of Standards and Science (KRISS), Daejeon 305-600, Korea}

\begin{abstract}
We have measured the angular correlation of the pair nucleons 
\np~and \nn~emitted from the nonmesonic 
weak decay (NMWD) of $^{12}_{\it{\Lambda}}$C~produced via the ($\pi^+$,$K^+$) 
reaction in coincidence measurement.
The ${\it{\Lambda}}p\to np$ and ${\it{\Lambda}}n\to nn$ modes were clearly
identified by measuring the back-to-back correlation of the emitted nucleon
pairs which is the characteristic of two-body kinematics.
From the measured nucleon pair numbers $N_{nn}$ and $N_{np}$, 
the ratio $\Gamma_n/\Gamma_p$~of the partial decay widths
$\Gamma_n$($\it{\Lambda}n\rightarrow nn$)
and $\Gamma_p$($\it{\Lambda}p\rightarrow np$)
of $^{12}_{\it{\Lambda}}$C~was extracted to be 
$0.51\pm0.13$(stat)$\pm 0.05$(syst); this result is almost free from 
the ambiguity due to the nuclear final state interaction and 3-body 
decay process, which were inherent in the previous results.
The obtained  $\Gamma_n/\Gamma_p$~ratio of $^{12}_{\it{\Lambda}}$C 
($p$-shell) is close to that of $^5_{\it{\Lambda}}$He ($s$-shell).
The results are consistent with those of recent theoretical 
calculations.
\end{abstract}

\end{frontmatter}

\section{Introduction}

A $\it{\Lambda}$~bound in a nucleus decays via either mesonic weak decay,
$\it{\Lambda}$ $\to N\pi$, or nonmesonic weak decay (NMWD),
$\it{\Lambda} N \to n N$. Mesonic decay is essentially similar to 
free $\it{\Lambda}$ decay and has been studied in detail. 
In a nucleus a $\it{\Lambda}$ can decay via an interaction with
a neighbor nucleon, either a proton
($\it{\Lambda}p\to np; \Gamma_p$, proton-induced NMWD) or a neutron
($\it{\Lambda}n\to nn; \Gamma_n$, neutron-induced NMWD); this is referred
to as a one-nucleon-induced (1N) NMWD process. 
The two-nucleon induced (2N) process
($\it{\Lambda}$$NN \rightarrow n N N$; $\Gamma_{2N}$), which is 
another NMWD mode, has been 
predicted theoretically; however, it has not yet been verified 
experimentally.   
The NMWD process has attracted considerable attention since it provides
the only practical means of studying the 
strangeness-changing baryonic weak interaction at present~\cite{Alb02}.  

Since the discovery of hypernuclei, one of the primary concerns in the 
study of NMWD of $\it{\Lambda}$~hypernuclei has been the relative strength of 
the two channels of 1N NMWD-the ratio of the decay widths (\Gn/\Gp).
Experimental ratios over the broad mass
range of $\it{\Lambda}$~hypernuclei 
are consistently greater than unity, thereby indicating the 
dominance of the neutron channel; on the other hand, the theoretical 
ratios are only one tenth of unity. The predominance
of the proton channel in the theoretical prediction is due to
the fact that the tensor term 
of one-pion exchange (OPE) contributing only to the proton
channel is very high, and therefore, the contributions of other meson 
exchange terms become minor corrections. 
The OPE contribution is included in all model calculations for the 
long range contribution; hence, the $\Gamma_n/\Gamma_p$~ratios 
have remained at around 0.1.
This intriguing problem is referred to as the 
$\Gamma_n/\Gamma_p$~puzzle. 

In order to reduce the predominance of the proton-induced NMWD 
of the OPE model, many theoretical models for short range 
contribution, such as 
the heavy meson exchange (HME) models and the direct quark (DQ) model 
have been studied. All these models have been unsuccessful in 
increasing the $\Gamma_n/\Gamma_p$~ratio significantly until recently. 
However, in recent important development,  
the incorrect sign in kaon exchange amplitudes was identified; 
the correction significantly increased the $\Gamma_n/\Gamma_p$~values~of 
these models \cite{Sas00}.
Thus, the current theoretical values of the $\Gamma_n$/$\Gamma_p$ 
ratio have increased to 0.34-0.70 for $^{5}_{\it{\Lambda}}$He 
and 0.29-0.53 for $^{12}_{\it{\Lambda}}$C~\cite{Sas00,Jid01,Par01}.

Until recently experimental $\Gamma_n/\Gamma_p$~values were 
extracted by fitting intranuclear cascade 
(INC) calculation yields to the experimental proton yields spectra.
Almost all the experimental $\Gamma_n/\Gamma_p$ ratios have been close to 
or greater than unity~\cite{Alb02}. 
However, recent high-quality neutron yield spectra 
obtained by our group have made it possible to derive
the $\Gamma_n/\Gamma_p$~ratio directly by comparing 
the neutron yields to the corresponding proton yields;
the \Gnp~ratios obtained by this method are $\sim$0.5 
for $^{12}_{\it{\Lambda}}$C~\cite{Kim03,Oka04} and $\sim$0.6 for
$^5_{\it{\Lambda}}$He~\cite{Oka04} agreeing well with the
recent theoretical values.
These ratios are significantly smaller than unity. They are the first
experimental indications of the proton channel
dominance ($\it{\Lambda} p \to n p$) in the NMWD.
 
It appears that the discrepancy 
between the experimental and theoretical values of the \Gnp~ratios
of NMWD has been removed.
However, the $\Gamma_n/\Gamma_p$~ratios of \He5lam~and
\Clam~derived by comparing the nucleon yields still
contain the ambiguity due to the possible contributions of 2N NMWD
and the final state interaction (FSI) effects 
on the emitted nucleons. 
The current FSI model calculation which takes into account the 
2N NMWD and FSI can not reproduce the yield spectra~\cite{Oka04,Gar04}.
Therefore, the uncertainty of the ratios due to the ambiguity could not be 
estimated in such single nucleon  
(singles) measurements.   
In order to remove such ambiguity experimentally, the decay channel 
has to be explicitly identified for each  event. 
In order to determine each decay channel of 1N-induced NMWD
exclusively, we have performed a coincidence measurement of 
both  n+p ($np$ pair) and n+n ($nn$ pair) pair
nucleons from the NMWD of the hypernuclei, \He5lam~and \Clam. 
Since $\it{\Lambda}$N $\rightarrow$ NN is a two-body process, the outgoing
nucleon pairs will exhibit a clear back-to-back ($\textit {bb}$) 
correlation with regards to 
their opening angle, and their energy sum distribution would 
exhibit a peak at
around the decay Q value, $\sim$155 MeV, although they are broadened 
due to nuclear medium effects. On the other hand, in 2N NMWD, 
three nucleons are produced  in the final state
and a pair of nucleons among them would exhibit  
neither the  $\textit{bb}$ feature in 
the opening angle nor the energy sum peak at the Q value.
The nucleon from 1N NMWD that suffered intense FSI 
on the way out of the residual
nucleus would be deflected from the original momentum 
direction and lose kinetic energy, 
thereby degrading the \bb kinematics characteristic of 1N NMWD.    
Therefore, by applying the event selection criteria of \bb kinematics, 
we can exclusively select 1N NMWD events 
suppressing possible contributions from the 2N process 
and the events that suffered serious FSI.

Recently we have reported the $\Gamma_n/\Gamma_p$~value of $s$-shell 
$^{5}_{\it{\Lambda}}$He that was determined in the two-nucleon coincidence 
measurement of the NMWD~\cite{Kan05}. By applying the \bb kinematical
condition for the opening angle, 
we obtained the pair number ratio of $nn$ to $np$ pairs of 1N NMWD, 
$N_{nn}/N_{np}$, to be 0.45$\pm$0.11$\pm$0.03. 
This ratio becomes essentially the \Gnp~ratio when FSI is weak 
as in \He5lam. 
The $s$-shell hypernucleus, $^5_{\Lambda}$He, was selected 
for the first exclusive coincidence measurement, 
since the FSI effect for such a light hypernucleus must be small. 
However, the unambiguous determination of \Gnp~for a $p$-shell hypernucleus
also is essential in the study of baryonic weak interaction 
since only the $p$-shell hypernuclei can provide the $p$-wave initial
state for the interaction while the $s$-shell \He5lam~does only 
that of $s$-wave.
The importance of the $p$-wave contribution in the NMWD in the finite
nucleus has been discussed in the reference~\cite{Ben92}.
In such respect, \Clam~has been extensively studied together with \He5lam. 
Exclusive determination is even more important for the $\Gamma_n/\Gamma_p$~ratio 
of heavier $p$-shell hypernuclei such as \Clam~since 
the nucleons propagating there will suffer more intense FSI. 
Although $^{12}_{\it{\Lambda}}$C~is
the most extensively studied $p$-shell $\it{\Lambda}$~hypernucleus,
no unambiguous $\Gamma_n/\Gamma_p$~ratio for it is available yet. 

The values of the asymmetry parameter $\alpha_{nm}$ of protons from
$\it{\Lambda} p \rightarrow n p$ NMWD of the polarized hypernuclei
were measured and reported for both $^5_{\it{\Lambda}}$He 
and \Clam~formed via the ($\pi^+$,$K^+$) reaction. 
Although small positive values of 0.24$\pm$0.22 and 
0.11$\pm$0.08$\pm$0.04~\cite{Aji00,Mar05} were observed
for $^5_{\it{\Lambda}}$He,  
a large negative value ($\alpha_{nm} \simeq -1.0\pm 0.4$) was 
reported for the $p$-shell hypernuclei \cite{Aji92}.
Authors indicated a possible explanation that 
the interaction mechanism of NMWD for short and long range 
could be different and therefore the initial
$p$-state contribution in \Clam~made the difference. 
The theoretical predictions for the asymmetry parameter 
for \He5lam~are about -0.7~\cite{Par01,Sas01}
from both the recent heavy meson exchange (HME) and the direct quark (DQ)
models. They significantly overestimate the magnitude of the parameter. 
The recent HME model prediction for $\alpha_{nm}$ of \Clam~is 0.36
that is again far off from the experimental value. 
The current difficulty in understanding
the decay asymmetries of $s$- and $p$-shell hypernuclei implies 
that an accurate
and unambiguous determination of $\Gamma_n/\Gamma_p$~ratios for both
$s$- and $p$-shell hypernuclei is essential for NMWD studies. 

In this paper we report the results of the two-nucleon observables 
measured in the coincidence experiment of the decay 
of $^{12}_{\it{\Lambda}}$C especially focusing 
on its $\Gamma_n/\Gamma_p$~ratio.

\section{Experimental setup and analysis}

\label{exp}
\begin{center}
\makeatletter
 \def\@captype{figure}
 \makeatother
  \epsfig{file=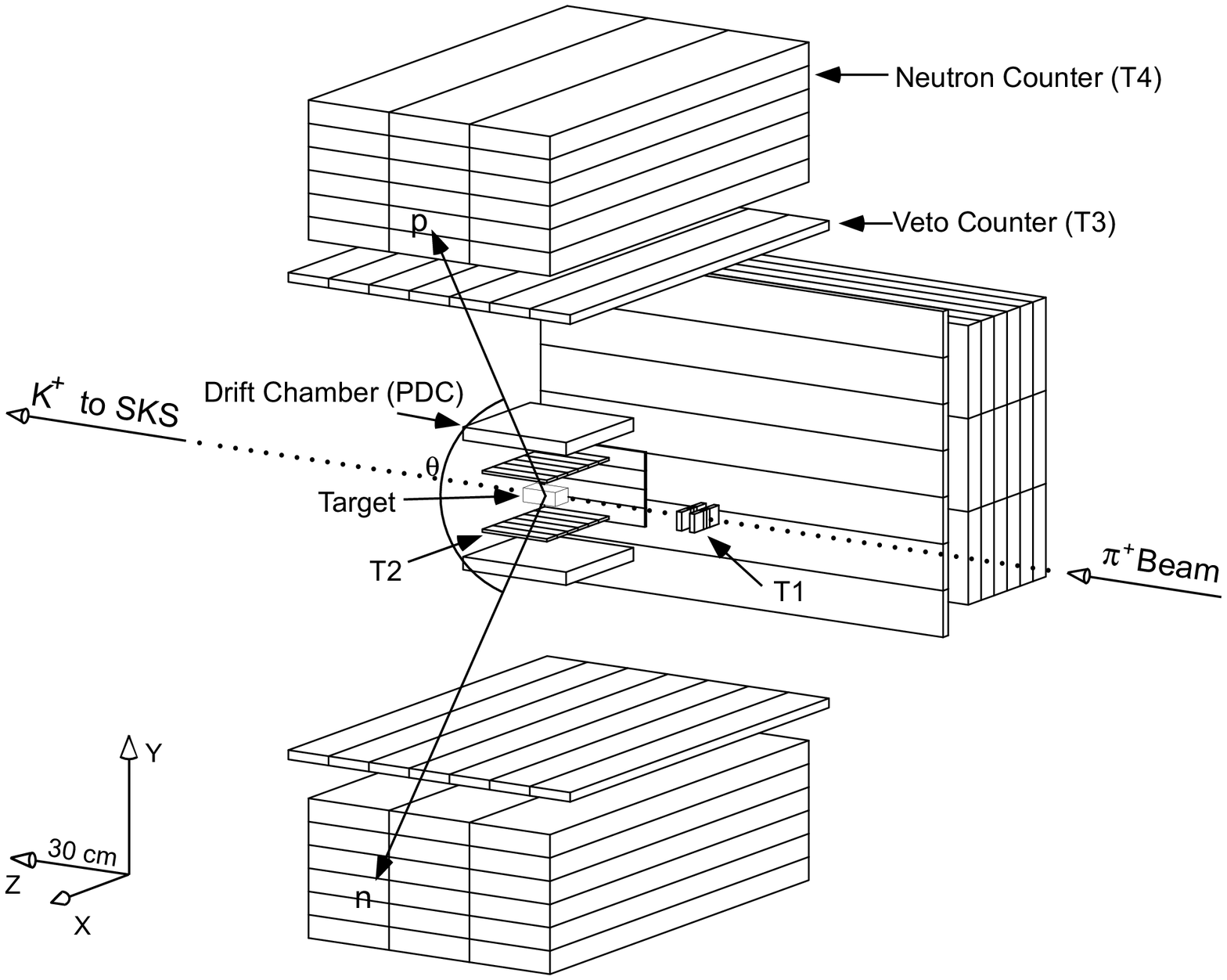,width=9.0cm}
 \caption{Schematic view of the decay-particle detection system is shown.
  Each of top and bottom counter sets comprise a fast timing counter (T2),
 a drift chamber (PDC), veto or stop timing
 counter (T3) and neutron counter arrays (T4). 
 The side set is similar; however, the PDC is absent.}
 \label{fig:e508setup}
\end{center}

The experiment was carried out at the K6 beam line of 12-GeV proton
synchrotron (PS) in High Energy Accelerator Research Organization (KEK). 
$\it{\Lambda}$ hypernuclei $^{12}_{\it{\Lambda}}$C were produced 
via the ($\pi^+$,$K^+$) reaction at the beam momentum of 1.05 GeV/c 
on the $^{12}$C target.
In order to improve the vertex resolution, we used thin segmented
plastic scintillators for an active $^{12}$C target. 
Fig.\ \ref{fig:e508setup} shows the schematic view 
of coincidence counters 
for detecting the decay product particles from
the weak decay of $^{12}_{\it{\Lambda}}$C. 
It comprises three sets of coincidence 
counter-two located at the top and bottom of the target, 
optimized for $\textit{bb}$
coincidence events and one at the side to allow the detection of 
non-back-to-back events. Each of top and bottom counter sets 
comprise a fast timing counter (T2), a drift chamber (PDC), 
veto or stop timing
counter (T3) and neutron counter arrays (T4). 
The side set is similar; however, the PDC is absent.

The inclusive $^{12}_{\it{\Lambda}}$C~excitation energy spectrum derived
from the momenta of incoming pion and the outgoing kaon is shown 
in Fig.\ 2(a). 
The two single particle $\it{\Lambda}$~states (1$s$ and 1$p$) are clearly 
identified. The 1$p$ $\it{\Lambda}$~state is located just 
above proton emission
threshold; it decays into  $^{11}_{\it{\Lambda}}$B~by emitting a proton. 
Fig.\ 2(b) and 2(c) show the excitation energy spectra in coincidence 
with the emitted pair nucleons $np$ and $nn$, respectively.
The vertical lines show the applied gates of the decay measurement for
ground state of $^{12}_{\it{\Lambda}}$C. 
The enhancement of the yield in the quasi-free
$\it{\Lambda}$~region of the $nn$ coincidence excitation energy spectrum 
as compared with that of the 
$np$ pairs occurs due to the emission of two neutron 
via the absorption
of $\pi^-$~from the quasi-free $\it{\Lambda}$ mesonic decay.  

\begin{center}
 \makeatletter
 \vspace{-2cm}
 \def\@captype{figure}
 \makeatother
  \vspace{15mm}
 \epsfig{file=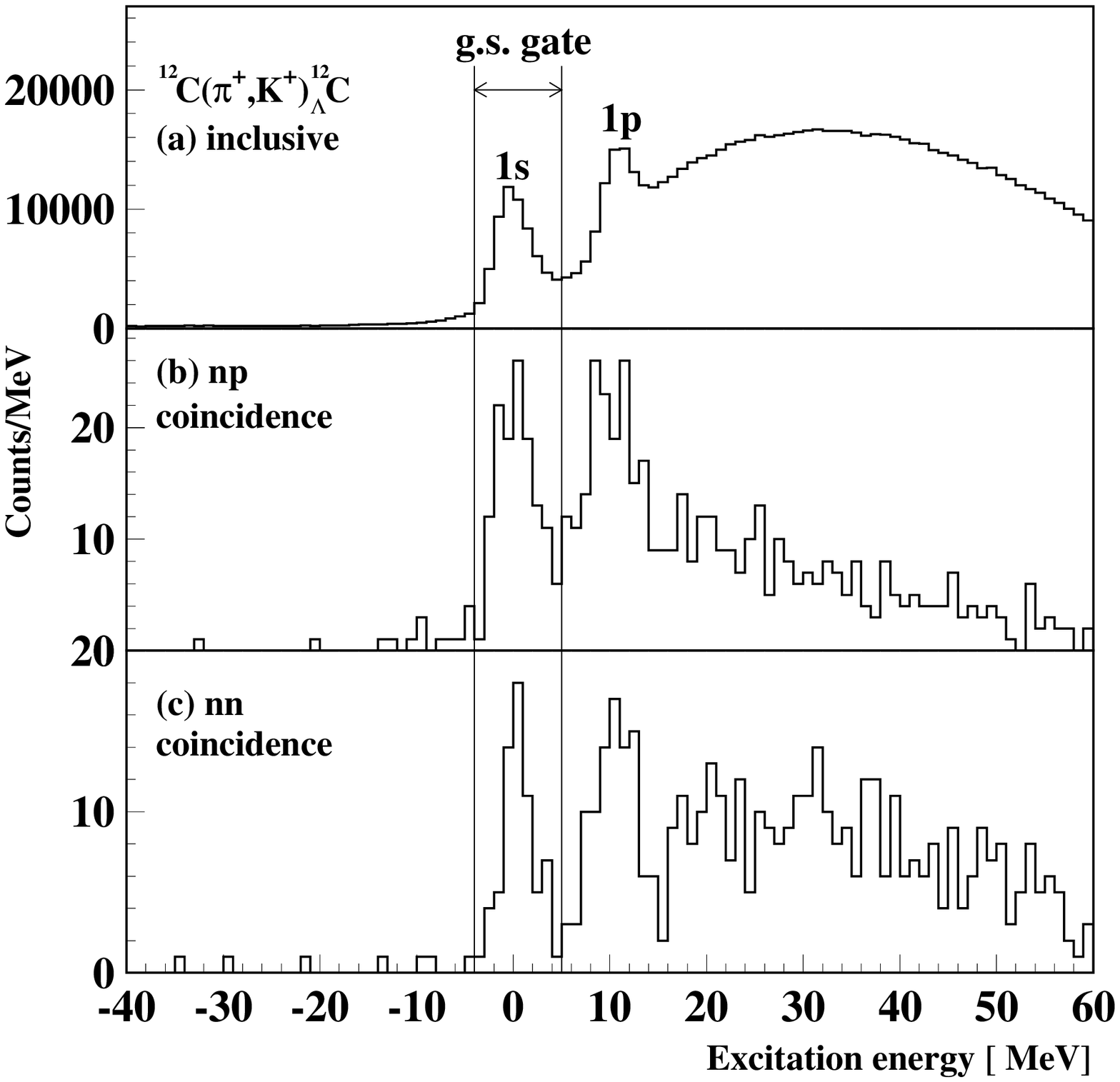,width=8.5cm,height=8.5cm}
 \caption{Excitation energy spectra of $^{12}_{\it{\Lambda}}$C,
         (a)inclusive, (b)with coincidence $np$ pairs,
         (c)with coincidence $nn$ pairs, respectively.
         (b) and (c) case, 30MeV energy cut condition was
         applied for both protons and neutrons.} 
 \label{fig:ex}
\end{center}

Neutral particles, $\gamma$ and neutrons, were measured using six layers
of 5 cm thick plastic scintillators (T4) vetoed with T3. 
The particle identification (PID) of the neutrons and $\gamma$ was 
achieved by using the flight time from the decay vertex 
to the fired T4 counter.
The neutron and $\gamma$ were well separated, and the accidental background
within the neutron gate was negligible. 
The charged particles were identified by combining 
$dE/dx$, ${\it TOF}_{23}$ and $E_T$.
$dE/dx$ denotes the energy loss per unit length measured 
by $T_{2}$; ${\it TOF}_{23}$, flight time between $T_{2}$
and $T_{3}$; and $E_T$, the total energy deposited in the sequentially
fired counters, $T_{2}$,~$T_{3}$, and $T_{4}$.
Protons are well separated from pions and deutrons. 
The kinetic energies of neutrons and protons were calculated from the 
flight times and the ranges, respectively. 
Proton energy was corrected for the energy  loss inside the target material
thanks to the good vertex resolution.
Further experimental details are provided in the previous report~\cite{Oka04}. 

\section{Results and Discussion}

Fig.\ \ref{fig:raw}(a) and \ref{fig:raw}(b) show the raw distribution
of $np$ and $nn$ pair yields $Y_{np}$(cos$\theta$) 
and $Y_{nn}$(cos$\theta$) in the  opening angle $\theta$, respectively.
Only the pair events of which each nucleon has an energy greater 
than 30 MeV are counted.
The dominance of the \bb peaks shown in the raw spectra is enhanced
by the maximized acceptance for the coincidence back-to-back 
kinematic events.  
A total of 116 and 43 events were observed in the $\textit{bb}$ 
angular region of cos$\theta < -0.7$  for the $np$ and $nn$ pair yields, 
respectively.
The shoulder bump observed in the \bb region of the \nn pair can 
be understood as a combined result of a statistical fluctuation due to 
the limited statistics of \nn pair
events (total 43 counts only in the \bb region) and the binning 
effect of neutron z-position analysis
(refer to Fig.\ \ref{fig:e508setup}). 
It was revealed that the background nucleon pairs produced through
the absorption of $\pi^{-}$ from the mesonic decay of
$^{12}_{\it{\Lambda}}$C~in the materials around the target were insignificant
unlike the situation in the case of $^{5}_{\it{\Lambda}}$He~\cite{Kan05}.
The angular resolutions for the \np and \nn pairs were estimated to be 
$\sigma_{\cos(\theta_{np})} = 0.018$ and 
$\sigma_{\cos(\theta_{nn})} = 0.029$ at cos$\theta = -0.9$, respectively.
\label{sec:discussion}
\begin{center}
 \makeatletter
 \def\@captype{figure}
 \makeatother
\vspace{-15mm}
 \epsfig{file=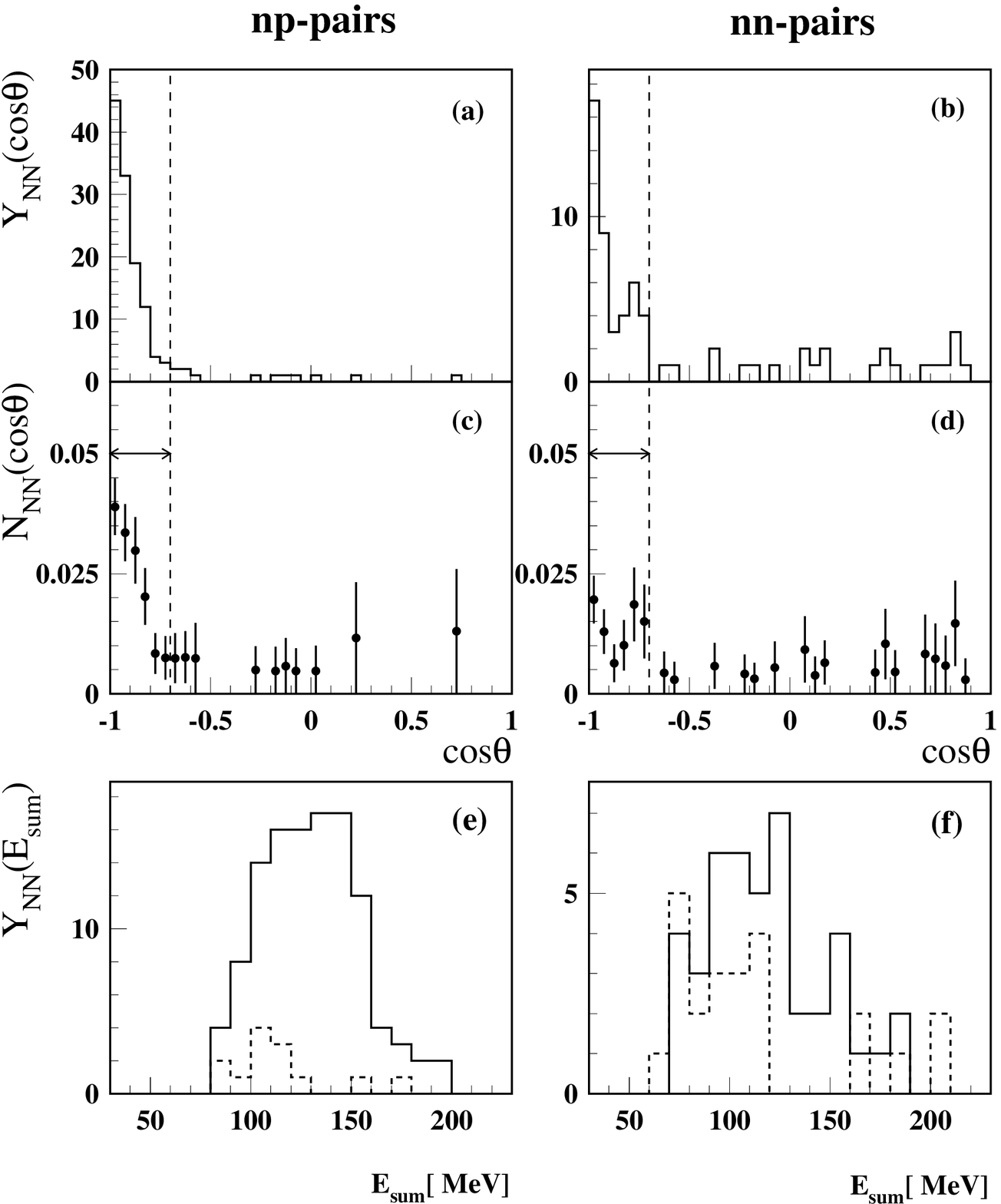,width=10.0cm}
 \caption{The distributions of $np$ and $nn$ pairs are shown in the left
and right figures, respectively. Top figures are the raw opening angle
distribution, while middle ones are the angular correlations for the
normalized pair numbers per NMWD for full solid angle and unit efficiency.
The solid line of bottom ones are the pair yields in the energy sum of the two nucleons
from the $\textit{bb}$ events indicated as the arrowed region,
cos$\theta < -0.7$. Totals of 43 and 116 events were observed for
$nn$ and $np$ pairs in the $\textit{bb}$ region, respectively.
Dashed lines in the bottom figures (e) and (f) indicate the energy sum spectra 
of the pair events in the non-\bb region. 
The energy threshold was 30 MeV for both protons and neutrons.} 
 \label{fig:raw}
\end{center}
Fig.\ \ref{fig:raw}(c) and \ref{fig:raw}(d) show the angular
correlation of pair nucleons, namely the normalized pair yields per NMWD,
$N_{NN}$(cos$\theta$).
$N_{NN}$(cos$\theta$) can be expressed 
as $N_{nn(p)}$(cos$\theta$)$=Y_{nn(p)}$(cos$\theta$)$/(Y_{nm}\cdot\epsilon_{nn(p)})$,
where $Y_{nm}$ and $\epsilon_{nn(p)}$ are the total number of NMWD observed
and the overall efficiency for detecting two nucleons from the \nn and \np 
pairs in coincidence including the detector acceptance.  
$Y_{nm}$ is obtained by multiplying the number of hypernuclei produced
in the ground state of $^{12}_{\Lambda}C$ with the NMWD branching ratio~$b_{nm}$,
which is defined as $b_{nm} = 1-b_{\pi^-} - b_{\pi^0}$,
where $b_{\pi^-}$ and $b_{\pi^0}$ denote the $\pi^-$ and 
$\pi^0$ branching ratios, respectively.
The accuracy of the $b_{nm}$ value of 0.768$\pm$0.012$\pm$0.005 
has been improved drastically due to the accurate measurement 
of $b_{\pi^0}$ in the present experiment~\cite{Oka05}
in comparison with the previous one~\cite{Sak91}. 
This accurate $b_{nm}$~value made it possible to normalize 
the pair yields per NMWD without introducing significant 
systematic errors.
The measured $N_{NN}$ angular correlation of $nn$ and $np$ pairs
could now be directly compared with those obtained from the FSI model 
calculation~\cite{Gar04}.
$\epsilon_{NN}$(cos$\theta$) was calculated with energy dependent
event-by-event simulations, and it was tested to be in good agreement
with the data~\cite{Kan04}. The neutron detection efficiency was calculated
by the Monte-Carlo simulation code-the modified DEMONS code-which
is applicable to a multi-element neutron detector and has been tested
for producing various experimental data well~\cite{Kim03}.
The angular correlation of $np$ pair clearly exhibits a peak at
cos$\theta$ $\simeq-1$, thereby indicating 
the $\textit{bb}$ emission of two
nucleons, which is the signature of two-body decay final state.
The \bb kinematic regions defined as cos$\theta < -0.7$
essentially include all the events in the \bb kinematics peak. 
However, although dominance of \bb kinematics events is 
degraded to a certain extent in the \nn pairs, the \bb events 
continue to be a major contribution. 
The dominance of the \bb kinematic events observed in both \np and 
\nn pair confirms that 1N NMWD, $\it{\Lambda}p\to np$ and
$\it{\Lambda}n\to nn$, is the main NMWD mode of \Clam.
In non-\bb region, we observe pair events more or less uniformly 
distributed over all angle in cos$\theta$ whose total number reaches
$\sim$40 \% of the total pair numbers as shown in Table 1.
Table~\ref{tab:pair_number} shows the pair numbers in each opening
angle region. 
We consider the uniform distribution event extended even in \bb region.
Therefore, we will
subtract the corresponding component of the uniform distribution events
in \bb region 
from the pair numbers $N_{NN}$ for the later discussion of 1N NMWD.   

\begin{table}
\begin{minipage}{\textwidth}
\begin{footnotesize}
\begin{center}
\caption{The yields of coincidence events $Y_{np}$, $Y_{nn}$ and 
$Y_{pp}$ and the normalized pair numbers per NMWD $N_{np}$ ,$N_{nn}$ 
and $N_{pp}$ are shown in the opening angle regions 
$\textit{bb}$ (back-to-back,cos$\theta<-0.7$) and $\textit{non-bb}$ 
(cos$\theta\geq-0.7$). $N_{NN}$ values listed are the numbers 
simply integrated over the corresponding angular regions.
The pair numbers after the subtraction of the uniform background in the 
$\textit{bb}$ region are also listed. Errors are statistical.}
\label{tab:pair_number}
\vspace{0.5cm}
\begin{tabular}{c|c c|c c|c c} \hline\hline
 Angular region & $Y_{np}$ &    $N_{np}$     &$Y_{nn}$ &     $N_{nn}$   &       $Y_{pp}$                             &    $N_{pp}$     \\ \hline
$\textit{bb}$       & 116      & 0.138$\pm$0.014 & 43      &0.083$\pm$0.014 &  8                                         & 0.005$\pm$0.002 \\   
$\textit{non-bb}$   & 12       &0.060$\pm$0.018  & 23      &0.083$\pm$0.020 & ~~0\footnote{due to the little acceptance} &                 \\ \hline 
$\textit{bb}$(b.g. subtracted) &    & 0.127$\pm$0.014  &   &  0.067$\pm$0.015  &                                          &0.005$\pm$0.002  \\  \hline \hline 
\end{tabular}
\end{center}
\end{footnotesize}
\end{minipage}
\end{table}

In Fig.\ \ref{fig:raw}(e) and \ref{fig:raw}(f), the spectra 
of the $np$ and $nn$ pair yields in the NMWD of $^{12}_{\it{\Lambda}}$C
in the $\textit{bb}$ region are shown in the energy sum of 
the two emitted nucleons. 
The rms energy-sum resolutions, $\sigma_{E_{sum}}$, for $nn$ and $np$
pairs were estimated to be 11 and 9 MeV for typical cases, 
such as 75 MeV nucleons, respectively.
The energy-sum spectra show that the pair yields are 
distributed over a broad energy region from 70$\sim$80 MeV to 
the Q value at around 155 MeV; on the other hand,
a few pair yields exist above the Q values. 
Those present above the Q values are considered to be due to the 
rapid deterioration in the neutron-energy resolution 
with an increase in the neutron energy. 
Since the energy sum spectrum of the two nucleons emitted in the 
two-body decay process, $\it{\Lambda} N \rightarrow NN$, would exhibit 
a peak at around the Q value, $\sim$155 MeV, the broad energy distribution
indicates that a large number of the \bb kinematics nucleons 
suffered an energy loss on the way out of the residual nucleus 
with a mass of $\sim 10$. On the other hand, a dominant narrow peak  
around Q value was clearly visible in the \np pair energy sum spectrum 
of $^5_{\it{\Lambda}}$He~\cite{Kan05}.  
Therefore even if we consider the pair numbers confined 
in the $\textit{bb}$
region only, they include a considerable number of events 
that suffered FSI. 
We observe a similar or even more energy degraded energy-sum spectrum 
in $nn$ pairs;
however, the statistics are considerably lower.
\vspace{-5mm} 
\begin{center}
 \makeatletter
 \def\@captype{figure}
 \makeatother
 \epsfig{file=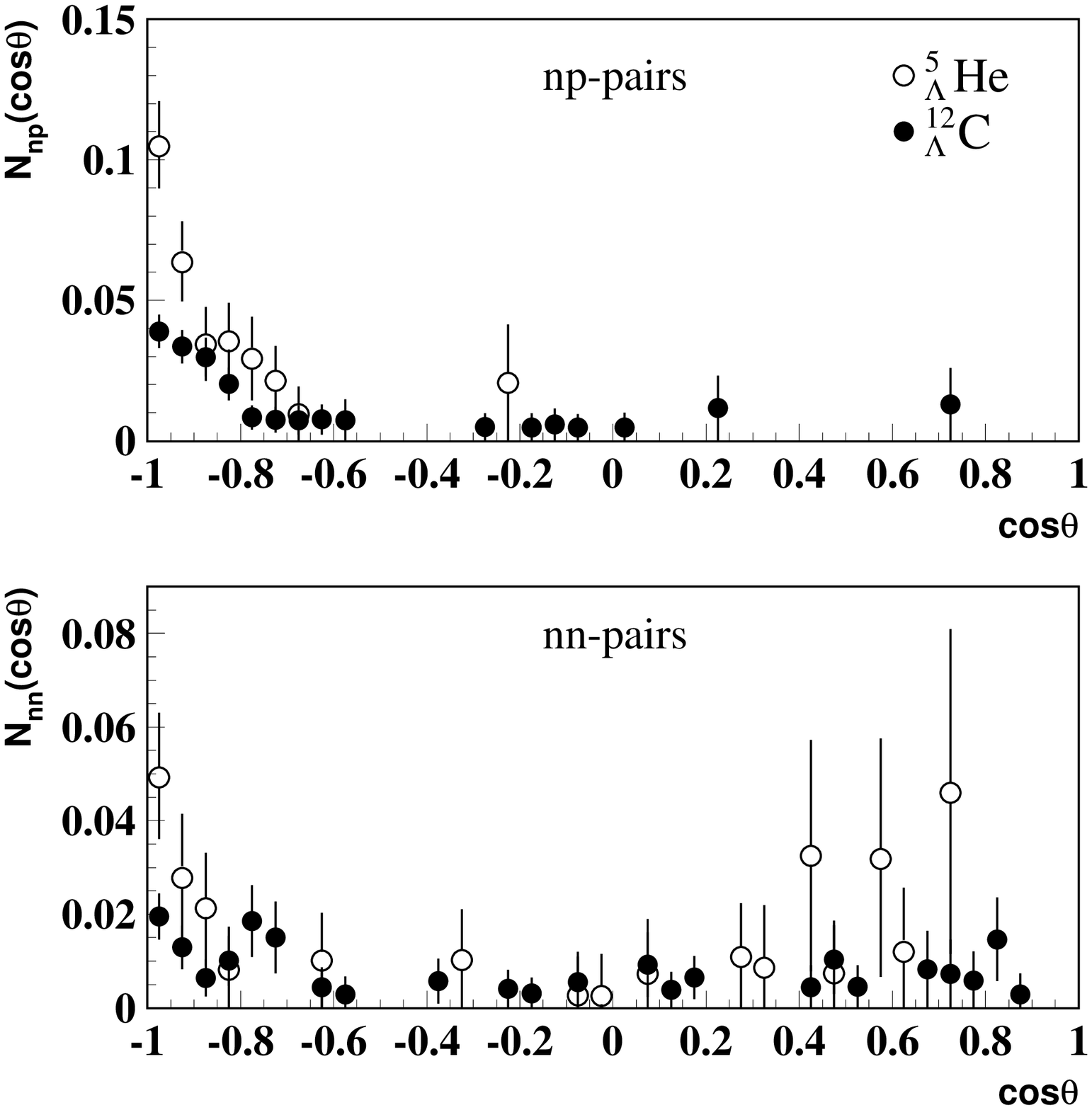,width=9.0cm}
 \caption{ $np$(top) and
           $nn$(bottom) pairs per NMWD  of $^{12}_{\it{\Lambda}}$C
           (filled circle) and $^{5}_{\it{\Lambda}}$He(open circle)
            are represented as a function of their opening angle.}
\label{fig:compare}
\end{center}
Fig.\ \ref{fig:compare} compares the angular 
correlations $N_{np}$(cos$\theta$) and $N_{nn}$(cos$\theta$) 
of \He5lam(open circle)~\cite{Kan05} and \Clam~(filled circle).
We observe dominant $\textit{bb}$ peaks in both $^{5}_{\it{\Lambda}}$He
and $^{12}_{\it{\Lambda}}$C; this indicates that the 1N NMWD is 
the major mechanism of NMWD. 
By comparing the angular correlations of $^{5}_{\it{\Lambda}}$He
 and $^{12}_{\it{\Lambda}}$C, we observe two clear features-broadening 
of the $\textit{bb}$ peak in angular correlations and quenching of the  
\bb pair numbers of $^{12}_{\it{\Lambda}}$C over those 
of  $^{5}_{\it{\Lambda}}$He in both $np$ and $nn$ pairs. 
The total  $\textit{bb}$ pair numbers of $np$ and $nn$ 
in $^{12}_{\it{\Lambda}}$C
are reduced in comparison with those of \He5lam~by 
$\sim$40\% and $\sim$20\%, respectively. 
The features can be understood in terms of the same physics 
responsible for the energy degradation in the 
$E_{sum}$($^{12}_{\it{\Lambda}}$C)
spectra (Fig.\ \ref{fig:raw}(e)) 
in comparison with those of $^{5}_{\it{\Lambda}}$He.  
By considering that 2N theoretical branching ratio does not vary 
considerably  
with the mass number, and therefore, the relative importance 
of 2N with respect to 1N would be
more or less the same for $^{5}_{\it{\Lambda}}$He 
and $^{12}_{\it{\Lambda}}$C~\cite{Gar04}, we infer that FSI is 
the major cause of broadening and quenching in the angular
correlation and the degradation in the energy-sum spectrum. 
This implies that the effect of FSI on the pair numbers, 
even in the $\textit{bb}$ region,
would not be negligible, and we must take it into account  
in order to extract  $\Gamma_{n}/\Gamma_{p}$ ratio from the pair number ratio. 
A significant portion ($\sim$40\%) of the pair number is observed in the 
non-back-to-back region as shown in Table~\ref{tab:pair_number}. 
The possible contributions in the non-back-to-back region 
are such as 2N(or multinucleon)-induced NMWD events, those that suffer 
strong FSI, $\pi^-$ absorption contamination  and random coincidence 
events etc. 
However, in $^{12}_{\it{\Lambda}}$C, the contributions 
from the last two cases 
are found to be almost negligible, and the first two are considered 
to be the major contributions in non-back-to-back region. 
These non-\bb events, whose strength reaches $\sim$40 percent 
of NMWD, are the sources of the ambiguity in the previous 
results of \Gnp~since they can not be distinguished from 1N NMWD 
in the singles measurement. 
However, we can reject them 
by considering only the coincidence pair numbers in the \bb region
in the derivation of \Gnp, thereby resulting the ratio
which is almost free from the ambiguity inherent in 
the previous results. 

Next, we would like to derive the $\Gamma_{n}/\Gamma_{p}$ ratio
from the pair number ratio, $N_{nn}/N_{np}$,~of the $\textit{bb}$
region. If neither $\Gamma_{2N}$ contribution nor FSI existed, the ratio of
$N_{nn}/N_{np}$ would become $\Gamma_{n}/\Gamma_{p}$.
However, as discussed, the effects of FSI still remain in the 
pair numbers of $^{12}_{\it{\Lambda}}$C. 
The most harmful effect of FSI is the channel crossover
such as that from \nn to \np or from \np to $nn$. 
It directly contaminates the pair number ratio. 
We observed 8 $pp$ pair coincidence events in the $\textit{bb}$ region, 
which 
are possible only via the FSI process. A neutron in the $np$ pair emitted at the vertex is
converted into a proton via the nuclear interaction on the way 
out of the nucleus, thereby registering as a $pp$ coincidence event.
Therefore, they represent the channel crossover information of FSI on the nucleons emitted.
Table~\ref{tab:pair_number} shows $N_{pp}$ yields-the normalized
$pp$ pair numbers per NMWD. 
We will utilize $N_{pp}$ for the correction 
of the FSI crossover effect 
on $\Gamma_n/\Gamma_p$~ratio. The pair numbers in the $\textit{bb}$ region
$N_{nn}$($\textit{bb}$), $N_{np}$($\textit{bb}$) 
and $N_{pp}$($\textit{bb}$)
can be expressed as 
\begin{center}
\begin{eqnarray}
  \label{eq:pairs_bb}
      N_{nn}(bb) =& r_{n}f_{n}^2p &+ r_{p}f_{n}g_{n}q    \nonumber \\ 
      N_{np}(bb) =& r_{p}f_{n}f_{p}p &+ 2r_{n}f_{n}g_{p}q \\
      N_{pp}(bb) =& r_{p}f_{p}g_{p}q &+ r_{n}g_{p}^2q^{'}    \nonumber 
\end{eqnarray} 
\end{center}
where $r_{n}(= \Gamma_{n}/\Gamma_{nm})$ and $r_{p}(= \Gamma_{p}/\Gamma_{nm})$
are the fractions of neutron and proton induced channels out of the NMWD. 
$f_{n(p)}$ is the loss factor of neutrons (protons) from the energy region of
interest due to FSI and $g_{n(p)}$ is the resulting crossover 
influx factor of
neutrons (protons) from protons (neutrons) channel. 
The first terms of $N_{nn}$ and $N_{np}$ represent the pair numbers emitted out
of the nucleus that maintain the channel starting 
at the initial weak vertex point.
The second terms represent the pair numbers that cross the channel 
and end in a channel different from the starting one. 
$p,q,$ and $q^{'}$ are the loss factors from the \bb angular region of the
original and channel crossing pairs; however,  
their actual values are not required in the following derivation.
Instead, they are replaced with the observed $pp$ pair numbers.
If there were no FSI effects, $f_{n(p)}$ and  $g_{n(p)}$ would be one 
and zero, respectively. We put $f_{n} =f_{p} =f$ and $g_{n}=g_{p} =g$ 
by considering the charge symmetry of a strong interaction. A similar shape
of proton and neutron spectrum obtained
from the present experiment \cite{Oka04} supports this approximation. 
A simple simulation reveals that the $g^{2}$ term in Eq.~\ref{eq:pairs_bb} 
is less than 1 percent of the noncrossing terms
and can therefore be neglected. 
Then, the second terms of $N_{nn}(bb)$ and  $N_{np}(bb)$ 
of Eq.~\ref{eq:pairs_bb} are replaced by  $N_{pp}(bb)$ and 
the $r_n/r_p$($=x=$\Gnp) ratio can be expressed in terms of 
pair numbers only as follows: 
\begin{eqnarray}
  \label{eq:gngp}
      2 N_{pp}(bb) x^2 - N_{np}(bb) x + N_{nn}(bb)-N_{pp}(bb)=0 \\ 
      \frac{\Gamma_n}{\Gamma_p} = x = 0.51 \pm 0.13\pm0.05.  \nonumber
\end{eqnarray} 

The $N_{NN}(bb)$ numbers are the integrated pair numbers in the \bb region
after the subtraction of the background, which is assumed to be uniformly distributed 
in all cos$\theta$ as we pointed out in the discussion of Fig. 3(c) and 3(d).  
The value of the constant background was determined by $\chi^2$ fitting. 
The ratio of the raw pair numbers simply integrated over the \bb region 
0.60$\pm$0.12 becomes 0.53$\pm$0.13 after the subtraction of the uniform background.
The obtained \Gnp~ratio $0.51 \pm 0.13\pm0.05$ in the Eq.~\ref{eq:gngp}~
is only 4 percent reduced from the pair number ratio 
$N_{nn}(bb)/N_{np}(bb)$. This is surprising considering that we observed the strong
FSI effect both in the energy-sum spectra and in the angular correlations. 
However, the small correction can be understood considering that 
$N_{nn}/N_{np}\sim 1/2$ and hence the crossover pair numbers 
from \nn to \np and from \np to \nn are almost balanced, and therefore,
the crossover-influx and -outflux due to FSI cancel each other.  
The systematic errors mainly originate from the uncertainties 
of the neutron detection efficiency (6\%) and the $pp$ pair acceptance (4\%). 
Many sources of uncertainty such as $b_{nm}, f_n$ and 
detection efficiencies are canceled out by taking the ratio~\cite{Kan05}. 

The present result determined with the pair nucleon number ratio 
$N_{nn}/N_{np}$ is in good agreement with the recent theoretical results based on 
heavy meson exchange (HME) model calculations~\cite{Jid01,Par01}.
It finally resolved the longstanding \Gnp~ratio puzzle 
without ambiguity by removing the effect of FSI self-consistently and by
rejecting non-\bb events applying \bb kinematic condition.
It is quite accidental that the present result is close to 
that of the previous singles measurement which were determined only by assuming
the nonexistence of 2N (or multi-nucleon induced) NMWD~\cite{Kim03}.
However, we have pointed out that there exist a significant amount
($\sim 40$\%~of the total) of non-\bb events, which probably are the candidates of them.    
The present $\Gamma_{n}/\Gamma_{p}$
ratio of $^{12}_{\it{\Lambda}}$C is in good agreement with that 
of the previous exclusive measurement of $^{5}_{\it{\Lambda}}$He 
($0.45\pm 0.11\pm 0.03$), which indicates that the NMWD mechanism 
of $s$-shell ($^{5}_{\it{\Lambda}}$He)
and $p$-shell ($^{12}_{\it{\Lambda}}$C) hypernuclei are rather similar. 
Though the \Gnp~ratio determined 
from the ratio of pair numbers $N_{nn}/N_{np}$ agrees well 
with those of the theoretical predictions~\cite{Par01,Gar04}, 
there remain discrepancies between the pair numbers $N_{NN}$
of our measurement and those of the theoretical calculation~\cite{Gar04}: 
the pair numbers for \bb events and non-\bb events, 
and the significant quenching of pair numbers of \Clam~compared 
with those of \He5lam.
For example, the observed pair number of \bb events is about half 
of the prediction while that of non-\bb events is only a fraction.

\section{Conclusion}

We have measured, for the first time, the angular correlation 
$N_{NN}$(cos$\theta$) of the pair nucleons \np and \nn
from the NMWD of \Clam~in coincidence measurement. 
We have exclusively identified the dominant 1N NMWD
process, $ {\it{\Lambda}}p\to np$ and ${\it{\Lambda}}n\to nn$, 
by requiring the \bb angular correlation of the two
emitted pair nucleons, which is the characteristic of two-body decay.
Then we have obtained the $\Gamma_n/\Gamma_p$ ratio from the ratio of 
pair numbers $N_{nn}/N_{np}$ of 1N NMWD in which most of the non-\bb events
are excluded and FSI effects have been corrected using the simultaneously
measured $pp$ pair events. Therefore, the present $\Gamma_n/\Gamma_p$
result is almost free from the ambiguities due to 2N(or multinucleon induced)
NMWD contribution and the FSI effects which were inherent in the previous
results obtained with singles spectra. 

The obtained  $\Gamma_{n}/\Gamma_{p}$ ratio of $0.51 \pm 0.13\pm0.05$
is in good agreement with the recent theoretical results calculated based on HME
models. This has finally resolved the long standing $\Gamma_{n}/\Gamma_{p}$
ratio puzzle unambiguously. This ratio is very close to that 
of the previous exclusive measurement of $^{5}_{\it{\Lambda}}$He,
thereby indicating the decay mechanisms of $s$-shell
($^{5}_{\it{\Lambda}}$He) and $p$-shell ($^{12}_{\it{\Lambda}}$C)
hypernuclei are rather similar. 
Moreover, we have obtained the pair number yields in the non-back-to-back 
kinematics region, which would provide information 
on the possible 2N NMWD contribution.
 
\section*{Acknowledgments}

We are grateful to Prof.\ K.\ Nakamura and KEK-PS staff for their support 
during our experiment and for ensuring the stable operation of KEK-PS.
The authors, MJK and HB, acknowledge the support provided by
KOSEF(R01-2005-000-10050-0) and KRF(2003-070-C00015).

\end{document}